\documentclass[reprint,amsmath,amssymb,aps,prl,superscriptaddress]{revtex4-2}
\usepackage{graphicx}
\usepackage{dcolumn}
\usepackage{bm}
\usepackage{soul}
\setstcolor{red}
\usepackage{physics}
\usepackage{xcolor}
\usepackage[colorlinks=true, linkcolor=blue, citecolor=blue, urlcolor=blue]{hyperref}
\usepackage{orcidlink}
\usepackage{lipsum}

\begin{document}

\preprint{APS/123-QED}

\title{Observation of Jones-Roberts solitons in a paraxial quantum fluid of light}

\author{Myrann Baker-Rasooli~\orcidlink{0000-0003-0969-6705}}
\affiliation{Laboratoire Kastler Brossel, Sorbonne Universit\'e, CNRS,
ENS-PSL Research University, Coll\`ege de France, 4 Place Jussieu, 75005 Paris, France}

\author{Tangui Aladjidi~\orcidlink{0000-0002-3109-9723}}
\affiliation{Laboratoire Kastler Brossel, Sorbonne Universit\'e, CNRS,
ENS-PSL Research University, Coll\`ege de France, 4 Place Jussieu, 75005 Paris, France}

\author{Nils A. Krause~\orcidlink{0009-0009-0754-7782}}
\affiliation{Department of Physics, University of Otago, Dunedin 9016, New Zealand}
\affiliation{Dodd-Walls Centre for Photonic and Quantum Technologies, Dunedin 9054, New Zealand}

\author{Ashton S. Bradley~\orcidlink{0000-0002-3027-3195}}
\email{ashton.bradley@otago.ac.nz}
\affiliation{Department of Physics, University of Otago, Dunedin 9016, New Zealand}
\affiliation{Dodd-Walls Centre for Photonic and Quantum Technologies, Dunedin 9054, New Zealand}

\author{Quentin Glorieux~\orcidlink{0000-0003-0903-0233}}
\email{quentin.glorieux@sorbonne-universite.fr}
\affiliation{Laboratoire Kastler Brossel, Sorbonne Universit\'e, CNRS,
ENS-PSL Research University, Coll\`ege de France, 4 Place Jussieu, 75005 Paris, France}


\begin{abstract}
We investigate the formation and dynamics of Jones-Roberts solitons in a smoothly inhomogeneous quantum fluid.
To do so, we create a superfluid of light using paraxial, near-resonant laser beam propagating through a hot rubidium vapor. 
We excite a bounded vortex-antivortex dipole in the superfluid and observe its transition to a rarefaction pulse and back, in agreement with the seminal predictions of Jones and Roberts. 
Employing an analogy with ray optics, we calculate the trajectory of the interacting vortices, deriving an effective refractive index from the inhomogeneous fluid density. 
Finally, we examine analytically and experimentally the superfluid velocity correlations, 
observing a transfer of coherence from incompressible to compressible velocity of the quantum fluid, a direct signature of the dynamical conversion between vortices and rarefaction pulse.
\end{abstract}

\maketitle

The dynamics of vortices and solitons have attracted significant interest because of their wide-ranging applications in physics.
Solitons are solitary waves that result from the balance between interaction energy and kinetic energy \cite{zakharov1972exact} and have been extensively studied in systems such as water waves \cite{Drazin1989}, optical fibers \cite{kivshar2003optical} and Bose-Einstein condensates (BECs) \cite{PhysRevLett.83.5198,denschlag2000generating,strecker2002formation,Frantzeskakis2010}.
However, in two- and three-dimensional systems, dark solitons are subject to instabilities, such as snaking instability, leading to decay into vortex rings or other structures, thereby limiting their lifetime \cite{staliunas_vortices_1994,anderson2000watching}. 
Considerable efforts have been made to stabilize these non-linear structures by understanding and controlling vortex dynamics, as the interactions between vortices and solitons are crucial for determining the overall stability of these solutions \cite{adhikari2002mean,theocharis2003ring,PhysRevA.77.045601,Barenghi2014}.
For example, hybrid soliton-vortex structures have been observed, where vortices can help mitigate soliton instabilities by creating robust configurations that persist over time \cite{komineas2003vortex,ginsberg2005observation}. 
Additionally, three-dimensional vortex solitons have been shown to be stable under non-local dipole-dipole interactions in BECs \cite{lashkin2009stable}.

Remarkably, a special class of solitons, known as Jones-Roberts solitons, has been predicted to be dynamically stable even in two and three dimensions \cite{jones_motions_1982,jones_motions_1986}.
Jones-Roberts solitons arise for a speed smaller than the speed of sound $c_s$, and a significant characteristic of these solutions is their ability to manifest in two distinct spatial shapes depending on their velocity.
At low speeds, they display a vortex-antivortex dipole bound state \cite{neely_observation_2010}, which transitions to a (vortex-free) rarefaction pulse as the velocity increases (below $c_s$). 
Experimentally, Jones-Roberts solitons, in particular rarefaction pulses, have been very challenging to observe in atomic BECs \cite{meyer_observation_2017}.

In recent years, paraxial fluids of light have emerged as an alternative platform to study quantum fluid behavior \cite{PhysRevLett.121.183604,glorieux2023hot,PhysRevA.98.023825,michel2018superfluid}. 
In these systems, the propagation of light through a nonlinear medium is described by an equation analogous to the Gross-Pitaevskii equation, which allows the study of interacting quantum vortex dynamics in an optical context \cite{baker2023turbulent,ferreira2024exploring,azam2022vortex,congy2024topological}.  
The system allows simultaneous high-resolution, non-destructive measurement of density and phase of the quantum fluid.
In this work, we use this paraxial fluid of light approach to investigate the dynamics of solitons and quantum vortices.
Interestingly, the velocity-dependent transition between the vortex-antivortex dipole and the rarefaction pulse solution, initially discovered theoretically by Jones and Roberts \cite{jones_motions_1982}, has been predicted to be observable in the presence of a smoothly inhomogeneous background~\cite{smirnov_dynamics_2012}.
To test this prediction, we created a 2D superfluid by propagating a near-resonant Gaussian laser beam through a hot rubidium vapor \cite{fontaine2018observation,piekarski2021measurement}.
We set up the initial state by imprinting two single vortices with opposite signs (vortex-antivortex dipole) into an inhomogeneous superfluid using a Spatial Light Modulator (SLM). 
From this out-of-equilibrium state, we let our system evolve freely and observed the spontaneous formation of a Jones-Roberts rarefaction pulse (vortex-free soliton) due to the changing local value of $c_s$.
At later time, the soliton velocity falls below the stability threshold and we observed a re-dissociation of the soliton into the vortex-antivortex dipole~\cite{neely_observation_2010}.
To fully characterize this dynamics we used an analogy with ray optics \cite{smirnov_dynamics_2012} and calculated the trajectory of the interacting vortices, deriving an effective refractive index from the inhomogeneous background density.
Finally, we compare the velocity autocorrelation function of vortex configurations and the rarefaction pulse~ \cite{bradley_spectral_2022,krause2024thermal} with analytical results, finding the incompressible vortex coherence is transferred to the compressible fraction of the quantum fluid as the rarefaction pulse forms.

\begin{figure}[ht!]
    \centering
    \includegraphics[width=0.99\linewidth]{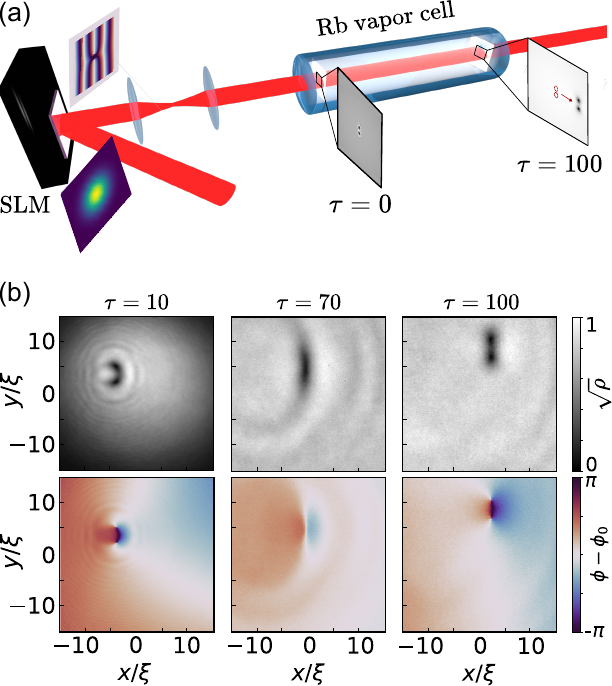}
    \caption{\textbf{Temporal evolution of a Jones-Roberts soliton.} \textbf{(a)} - Simplified setup. 
    A 780~nm laser beam is sent on an SLM and imaged at the input of the 20~cm-long Rb vapor cell. 
    The phase modulation leads to the creation of two counter-rotating vortices in the transverse plane. 
    The output plane of the nonlinear medium is imaged  after an interferometer (not shown).
     \textbf{(b)} - Top: experimental images of the field amplitude. 
     Bottom: associated phase. The phase $\phi_0$ measured in the absence of any vortices is removed. From left to right $\tau=10,70,100$.}
    \label{fig:setup}
\end{figure}
A paraxial fluid of light consists of a monochromatic laser beam propagating through a non-linear medium. 
In the paraxial approximation, the propagation equation of the laser electric field envelope $\mathcal{E}$ is isomorphic to the Gross-Pitaevskii equation describing the temporal evolution of the wavefunction for a weakly interacting quantum gas.
Each transverse plane at fixed z, is then a temporal snapshot of the evolution for an ideal 2D system and this reads as: 
\begin{equation}
    \label{eq:NLSE}
    i \pdv{\mathcal{E}(\mathbf{r}_\perp, z )}{z} \  =  \left[ -\frac{1}{2 n_0 k_0} \nabla_\perp^2 - \frac{k_0 \chi^{(3)}}{2 n_0} |\mathcal{E} (\mathbf{r}_\perp, z )|^2  \right]\mathcal{E} (\mathbf{r}_\perp, z ),
\end{equation}
where $k_0$ is the wavevector, $n_0$ is the linear refractive index given by $n_0 = \sqrt{1 + \text{Re}\chi^{(1)}}$, the $\nabla_\perp$ operator is defined as acting in the transverse $\mathbf{r}_\perp=(x,y)$ plane as a consequence of the paraxial approximation and the nonlinear term is proportional to the third-order susceptibility at the laser frequency $\chi^{(3)}$ times the laser intensity.
This last term induces an effective photon-photon interaction, and is set negative to ensure a stable superfluid with repulsive interactions.

The light field within the non-linear medium is not directly accessible experimentally, yet temporal evolution may be retrieved using an adimensional form of Eq.(\ref{eq:NLSE}).
This is done by incorporating the interaction term into a rescaled variable $\tau=~{L}/{z_{\text{NL}}}$, where $z_{\text{NL}}=\left[-k_0 \chi^{(3)} |\mathcal{E} (0, z )|^2/(2 n_0)\right]^{-1}$  is the characteristic nonlinear axial length and $L$ is the length of the non-linear medium \cite{bienaime_quantitative_2021}.
After re-scaling the transverse quantities ($\tilde{\textbf{r}}=\textbf{r}/{\xi}$, $\tilde\nabla_{\perp} = {\xi}\nabla_{\perp}$) by the transverse healing length ${\xi}= \sqrt{z_\text{NL}/k_0}$, one obtains for $\psi=\mathcal{E}/|\mathcal{E}|$:
\begin{equation}
    i\frac{\partial\psi }{\partial \tau}=
    \left(-\frac{1}{2}\tilde\nabla^2_{\perp}+{\mid}\psi{\mid}^2
    \right)\psi.
    \label{GPE_Adim}
\end{equation}
In this form, we show that the evolution of the system can be studied by tuning the ratio $\tau=~{L}/{z_{\text{NL}}}$ and in particular by tuning the laser intensity $|\mathcal{E} (0, z )|^2$.\\

In the experiment, we create a fluid of light by propagating a 780~nm laser set close to resonance (detuned by $\sim -10$GHz) of the $^{87}$Rb D2 line within a warm vapor cell ($\sim 150^\circ$C and $L=20$~cm) of rubidium which acts as a nonlinear medium (see Supplementary for details). 
As shown in Fig.~\ref{fig:setup}(a), we impose two localized counter-rotating phase circulations in the laser beam with an SLM.
The distance between the two singularities $\Delta\textbf{r}$ is fixed at the input of the medium and adjusted to keep the quantity $\Delta\textbf{r} / {\xi}$ constant while changing $\tau$ as their position from the center of the beam \cite{footnote1}.
The intensity and phase are recorded at the cell output using an off-axis interferometer \cite{PhaseUtils} and typical images are shown in Fig.\ref{fig:setup}(b) (top: amplitude, bottom: phase) for an increasing effective time $\tau$.
The effective times ($\tau=10,70,100$) are obtained by increasing the laser power, therefore reducing $z_{\text{NL}}$.

In Fig.\ref{fig:setup}(b), we show the direct formation of a Jones-Roberts soliton and its dynamics between the two distinct regimes of vortex dipole and rarefaction pulse \cite{jones_motions_1986}.
Since the background density is smoothly inhomogeneous, the dynamic of the vortices depends on the initial distance $\textbf{r}_{c}$ of the two-phase circulation to the center of the beam at $\textbf{r}=0$ \cite{smirnov_dynamics_2012}. 
We tune the initial state by setting this initial position off-centered ($x<0,y>0$). 
The amplitude and phase images at $\tau=10$ show the presence of two distinct vortices of opposite sign moving toward positive $x$ in the transverse plane. 
This is the low velocity limit (below the critical velocity of vortex-antivortex annihilation, $v_c$) for the Jones-Roberts bounded vortex dipole solution \cite{jones_motions_1986}.
At later time, this initial condition ensures that the vortices move toward regions of higher density (closer to the center), thus transforming into a vortex-free rarefaction pulse.
As $v$ approaches $v_c$ from below, the vortices move closer together, causing a depletion in the intermediate area and a loss of topological stability.
At time $\tau=70$, only one density minimum persists (which does not reach zero) and there is no phase singularity as shown in Fig.~\ref{fig:setup}(b).
Eventually, as the velocity decreases again, this rarefaction pulse transforms back to a bound vortex dipole at $\tau=100$.
These images provide a comprehensive observation of a Jones-Roberts soliton in a quantum fluid of light from the formation to the shape conversion and propagation.
In the following, we provide a detailed characterization of the observed behavior.


\begin{figure}[!t]
    \centering
    \includegraphics[width=0.95\linewidth]{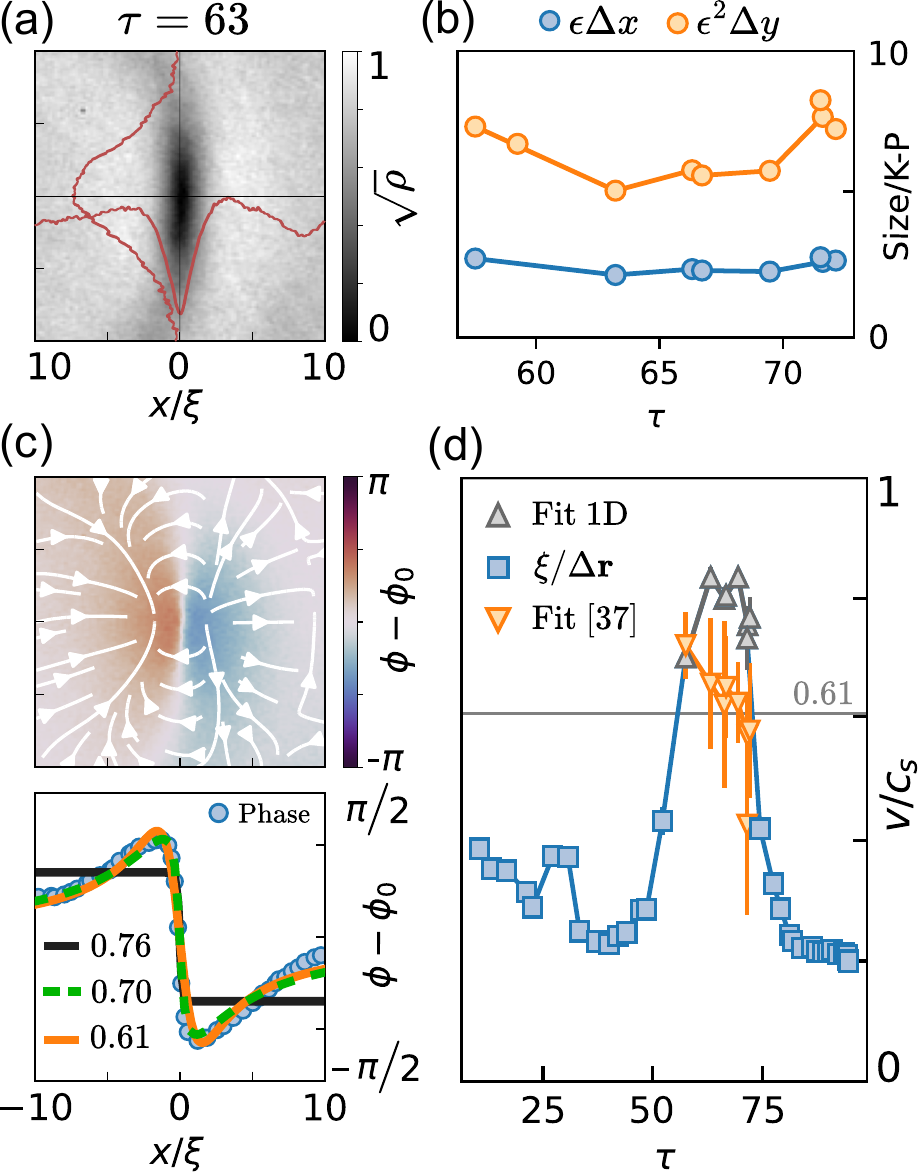}
    \caption{\textbf{Dipole stability as function of velocity.} 
     \textbf{(a)-(c)} Vortex-free rarefaction pulse observed at $\tau=63$. 
     \textbf{(a)} Amplitude image and profile along $x$ and $y$. 
     \textbf{(b)} Rarefaction pulse width $\Delta x$ and length $\Delta y$ compared to the K-P conditions versus time. 
     \textbf{(c)} Top: associated phase of (a) and velocity stream plots.
     Bottom: phase profile along $x/\xi$ in circles.
     The grey and orange solid lines show the fit results of the 1D solution Eq.~(\ref{soliton}) ($v/c_s=0.76$) and the asymptotic solution of \cite{tsuchiya_solitons_2008} ($v/c_s=0.61$), respectively.
     The green dashed line represents the analytical formula of \cite{berloff2004pade} at $v/c_s=0.7$ using Padé approximations.
     \textbf{(d)} - Velocity of the Jones-Roberts soliton as function of the evolution time. 
     The square markers give the value of $\xi/\Delta \textbf{r}$.
     The grey triangles represent the velocity extracted from the 1D solution Eq.~(\ref{soliton}). 
     The orange triangles shows the alternative results using the asymptotic solution of \cite{tsuchiya_solitons_2008}. 
     The grey line shows the theoretical value of $v/c_s=0.61$ of the transition between a bounded vortex dipole to vortex-free rarefaction pulse \cite{jones_motions_1986}.}
    \label{fig:velocity}
\end{figure}


To quantitatively compare our results with the analytical predictions of Jones and Roberts, we computed the velocity of the solitonic structure in units of $c_s$ as a function of evolution time $\tau$ and present the results in Fig.\ref{fig:velocity}(d). 
Velocities are calculated using different methods, depending on whether the vortices are well separated or have formed a rarefaction pulse (Fig.~\ref{fig:velocity}(c)). 
For widely separated vortices (low-velocity limit), the dipole speed is given by the velocity field generated by the respective other vortex, following the Biot-Savart law. 
This leads to the dipole speed $v=c_s\xi/\Delta\textbf{r}$, for the distance between the vortices $\Delta\textbf{r}$, similar to the hydrodynamics of two-dimensional point vortices \cite{lamb1932hydrodynamics, pismen_vortices_1999, fetter1966vortices,lucas2014sound}. 
In the intermediate velocity regime, there is no known analytical solution (to date) that describes a Jones-Roberts soliton once the vortices have merged \cite{krause2024thermal}.
To determine the velocity of the rarefaction pulse, we used two different approximations: the 1D soliton Eq.~(\ref{soliton}) and the 2D asymptotic solution (valid at high velocity) from \cite{tsuchiya_solitons_2008} to fit the phase in the vicinity of the phase jump, as illustrated in Fig.~\ref{fig:velocity}(c) (respectively black and orange line).
These two fits are compared with an approach based on Padé approximant (green dashed line) obtained at $v/c_s = 0.7$ \cite{berloff2004pade} that corresponds to our experimental data.
Following the assumption that the Padé approximant is the most precise method, we see that the 1D soliton approximation, with a narrow window around the phase jump, provides a better estimation of the rarefaction pulse velocity ($v/c_s=0.76$) than the 2D asymptotic solution ($v/c_s=0.61$).
This could be explained by the fact that the velocity range investigated in our experiment is far from the limit of $v\sim c_s$, where this 2D approximation \cite{tsuchiya_solitons_2008} is valid. 

\begin{figure}[!b]
    \centering
    \includegraphics[width=0.95\linewidth]{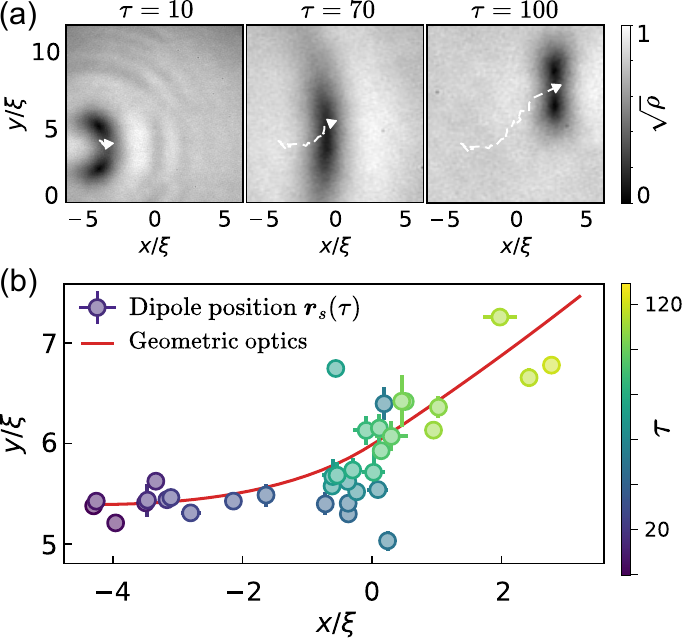}
    \caption{\textbf{Dipole trajectory in a smoothly inhomogeneous background.} \textbf{(a)} - Experimental images of the dipole amplitude for $\tau=10,70,100$. The dashed arrow shows the trajectory $\textbf{r}_s$  of the dipole/soliton.
     \textbf{(b)} - Position of the barycenter of the dipole/soliton in $\xi$ units for different time evolution $\tau$ given by the colorbar.
     The red curve shows the result of the geometrical optic equation (\ref{geom_optic}) solved numerically.}
    \label{fig:trajectory}
\end{figure}

We plot the velocity estimated by these two methods on Fig.\ref{fig:velocity}(d). 
The 1D fitting method slightly overestimates the value of $v/c_s$, while the method from \cite{tsuchiya_solitons_2008} underestimates it and is highly sensitive to noise, as evidenced by the error bars. 
By combining both methods, we conclude that the transition between the vortex dipole and the vortex-free rarefaction pulse takes place for $v\simeq v_c> 0.61c_s$, as expected for the critical velocity derived in \cite{jones_motions_1982, jones_motions_1986}, where the 2D solutions of Jones and Roberts lose vorticity (horizontal grey line). 
At $\tau > 52$, we see a transition from a vortex dipole ($v<v_c$) to a vortex-free rarefaction pulse ($v>v_c$) and back to a dipole at $\tau>72$, in precise quantitative agreement with the prediction of \cite{jones_motions_1982,jones_motions_1986, smirnov_dynamics_2012}.


Although no general analytical form of the rarefaction pulse is known, the density profile of the Jones-Roberts soliton for $v>v_c$ satisfies the Kadomtsev-Petviashvili (K-P) condition \cite{kadomtsev1970stability, jones_motions_1986, meyer_observation_2017}.
This condition states that the soliton width $\Delta x$ and length $\Delta y$, respectively, are scaling with $\xi/\epsilon$ and $\xi/\epsilon^2$, where $\epsilon = \sqrt{2}\sqrt{1 - v/c_s}$ in the low velocity approximation.
This scaling is verified by Fig.~\ref{fig:velocity}(b), wherein $\epsilon \Delta x$ and $\epsilon^2 \Delta y$ are plotted as functions of $\tau$, both of which remain relatively constant over time.


Moreover, as proposed theoretically in \cite{smirnov_dynamics_2012}, we determine the trajectory $r_s(\tau)$, along which a localized two-dimensional dark soliton moves with velocity $v/c_s$ using an analogy to ray optics. 
In a smoothly inhomogeneous background, the propagation of the dark soliton is given by standard ray optics law with an effective refractive index given by:
\begin{equation}
    \nu(\textbf{r}) = \sqrt{\rho_0(\textbf{r})}\frac{a(\textbf{r})}{\mathcal{E}_0 \rho_0(\textbf{r}_{s_0})}\textrm{sinh}\left( \frac{\mathcal{E}_0 \rho_0(\textbf{r}_{s_0})}{a(\textbf{r}) \rho_0(\textbf{r})} \right),
    \label{refractive_index}
\end{equation}
with $a(\textbf{r}) = 2\pi + (2\pi/3)\textrm{exp}\left[-\left(\mathcal{E}_0 \rho_0(\textbf{r}_{s_0})/(9.8 \rho_0(\textbf{r})) \right)^2 \right]$,
where $\mathcal{E}_0$ and $\rho_0(\textbf{r}_{s_0})$ are the normalized energy of the dipole and the density of the undisturbed background at the initial position of the dipole.
The trajectory is obtained by solving the geometrical-optics equation
\begin{equation}
    \frac{\partial^2\textbf{r}_s}{\partial \tau^2} = \frac{1}{2}\boldsymbol{\nabla}(\nu^2)|_{\textbf{r} = \textbf{r}_s},
    \label{geom_optic}
\end{equation}
with the initial condition
$   \frac{\partial \textbf{r}_{s_0}}{\partial \tau} = \nu(\textbf{r}_{s_0})\frac{\dot{\textbf{r}}_{s_0}}{|\dot{\textbf{r}}_{s_0}|},
$
where $\textbf{r}_{s_0}$ and $\dot{\textbf{r}}_{s_0}$ are the initial position and velocity of the dipole.

Figure \ref{fig:trajectory}(a) presents amplitude images corresponding to $\tau=10,70,100$, along with the experimental trajectory of the solitonic structure indicated by the white line.
Figure \ref{fig:trajectory}(b) illustrates the temporal evolution of the soliton/dipole center's position, demonstrating good agreement with the ray optics predictions (in red), thereby validating the theoretical model proposed in \cite{smirnov_dynamics_2012}.

\begin{figure}[!t]
    \centering
    \includegraphics[width=0.98\linewidth]{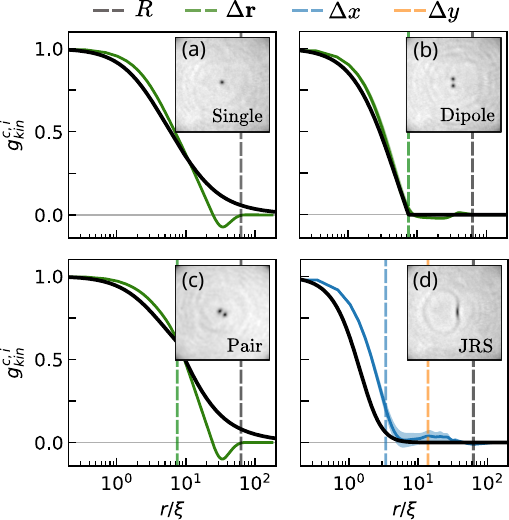}
    \caption{\textbf{Velocity correlations.} Experimental velocity two-point correlation function in $\xi$ unit.
    The angle-average velocity two-point correlator is calculated for (a) a single vortex, (b) a vortex dipole, (c) a vortex pair and (d) a vortex-free soliton using \cite{bradley_spectral_2022}, with quantum fluid densities inset. 
    The blue and green solid curve shows the compressible and incompressible part of the kinetic energy, respectively, with their associated analytical form in dark solid lines.
    Various color dashed lines represent $\Delta x$ (blue) and $\Delta y$ (orange) at $v/c_s=0.77$, the dipole inter-vortex distance $\Delta r = 7.5\xi$ (green) and the fluid radius $R=~63\xi$ (black).}
    \label{fig:corr}
\end{figure}

Finally, direct access to the fluid phase, shown in Fig.~\ref{fig:setup}(b), allows a measurement of the fluid velocity field, given by $\boldsymbol{v}^{tot}(\boldsymbol{r}) \propto \boldsymbol{\nabla}_{\perp}\phi(\boldsymbol{r})$ (see Supplementary for details).
We introduce the density-weighted velocity, given by $\boldsymbol{u}^{tot}(\boldsymbol{r})\equiv\sqrt{\rho(\boldsymbol{r}})\boldsymbol{v}^{tot}(\boldsymbol{r})$, where $\rho(\boldsymbol{r})$ is the light intensity.
We then use the Helmholtz decomposition \cite{baker2023turbulent,panico2023onset,PhysRevA.80.023618} to identify the divergent (compressible, $\boldsymbol{u}^c(\boldsymbol{r})$) and rotational (incompressible, $\boldsymbol{u}^i(\boldsymbol{r})$) parts of $ \boldsymbol{u}^{tot}(\boldsymbol{r})\equiv\boldsymbol{u}^c(\boldsymbol{r}) + \boldsymbol{u}^i(\boldsymbol{r})$.
The compressible part is associated to the contribution of acoustic waves and it can be subtracted from the total density weighted velocity to obtain the incompressible field associated with vortices (see Appendix A for details).
We then compute the velocity power spectrum and the two-point correlation function using spectral analysis for compressible fluids \cite{bradley_spectral_2022}.
The velocity power spectrum is given by
\begin{equation}
    E^{c,i}_{kin}(k) = \frac{mk}{4\pi}\int d^2r\,J_0(k|\boldsymbol{r}|)C[\boldsymbol{u}^{c,i}, \boldsymbol{u}^{c,i}](\boldsymbol{r}),
    \label{e_spec}
\end{equation}
with $C[\boldsymbol{u}^{c,i}, \boldsymbol{u}^{c,i}](r)$ being the two-point correlation in position and $J_0$ is the zero order Bessel function.
The system-averaged two-point velocity correlation is then given by
\begin{equation}
    g^{c,i}_{kin}(r) = \int dk J_0(kr)E^{c,i}_{kin}(k)/\int dk E^{c,i}_{kin}(k),
    \label{cor_spec}
\end{equation}
providing a measure of velocity coherence.

In Fig.~\ref{fig:corr}, we show the system averaged two-point velocity correlation function for (a) a single vortex, (b) a vortex dipole (opposite sign), (c) a vortex pair (same sign), and (d) a vortex-free rarefaction pulse. 
Figs. (b) and (d) correspond to the Jones-Roberts soliton, while (a) and (c) are basic configurations presented for reference. 
Each result is obtained from the experimental kinetic energy spectra (see Supplementary for details) and averaged over four images with a $\pi/2$ rotation.
We compare these results with our analytical predictions detailed in the End-Matter.
For (a-c), when vortices are present, we show the correlations between the incompressible component of the velocity field (green curves).
When the net vorticity is non-zero, there is a rotating flow with long-range correlations [Fig.~\ref{fig:corr}(a) and (c)]. 
Due to the finite size of the system (not present in the analytical model), these correlations become negative around the system size $R$ as the two-point correlation is dominated by counterflow at this scale.
For the vortex pair, the distance between the 2 vortex cores $\Delta r$ sets a length-scale where correlations are enhanced above the single vortex,  well described by our analytical model plotted in black line using Eq.~(\ref{gp}).
In stark contrast, in the case of the Jones-Roberts vortex dipole (Fig.~\ref{fig:corr}(b)), the correlations quickly decay to zero at the dipole separation distance, reflecting the low energy associated with this configuration, and in agreement with Eq.~(\ref{gd}).
Finally, in the rarefaction pulse regime, we observe a similar correlation scale to the vortex dipole, now seen in the compressible velocity correlation function since no more vortices are present, Fig.~\ref{fig:corr}(d). The vertical and horizontal length-scales of the rarefaction pulse are shown for comparison.


In this work, we experimentally investigated and characterized the temporal evolution of a vortex dipole in a smoothly inhomogeneous background within a two-dimensional quantum fluid of light. 
This platform allows to precisely characterize the spontaneous transition of a vortex dipole into a localized solitary wave (rarefaction pulse), and vice versa, as predicted for the Jones-Roberts soliton class. 
Additionally, by drawing an analogy with geometrical optics, we verified that the trajectory of the vortex dipole could be predicted by deriving an effective refractive index from the background undisturbed density. 
Furthermore, by extracting the compressible and incompressible components of the density-weighted velocity, and applying a high-resolution spectral analysis method, we reconstructed the two-point correlation function of our system with high resolution.
We explored the behavior of a single vortex, a vortex pair/dipole, and a Jones-Roberts rarefaction pulse, and compared the results with our analytical predictions. 
These findings show a clear observation of Jones-Roberts soliton in a fluid of light and deepen our understanding of contra-rotating vortices and soliton dynamics.
Our work opens a new perspective for studying macroscopic vortex behavior, such as vortex turbulence and its decay into wave turbulence \cite{PhysRevA.102.043318,PhysRevE.106.014205,PhysRevLett.128.224501}, and, more broadly, out-of-equilibrium quantum fluid physics.


The authors acknowledge insightful discussions with Iacopo Carusotto, Nicolas Pavloff, Pierre-Élie Larré, Thibault Congy, Clara Piekarski, Killian Guerrero, Thibault Bourgeois and Alberto Bramati. We thank financial support from the Agence Nationale de la Recherche (NC for grant ANR-19-CE30-0028-01 CONFOCAL and QG for grant ANR-21-CE47-0009 Quantum-SOPHA). QG is member of the Institut Universitaire de France (IUF).

\bibliography{biblio}

\begin{thebibliography}{51}
\bibitem [{foo()}]{footnote1}%
  \BibitemOpen
  \href@noop {} {}\bibinfo {note} {When renormalizing the transverse plane ${\xi}$ is the temporally averaged (i.e. along~$z$) healing length that accounts for losses ($\sim$30\%) and expansion during propagation.}\BibitemShut {Stop}%
\end{thebibliography}

\clearpage
\onecolumngrid
\section*{End Matter}
\twocolumngrid
\textit{Appendix A: Velocity decomposition.} 
From the phase map, we reconstruct the total velocity field, defined as $\boldsymbol{v^{tot}}(\boldsymbol{r}) = \boldsymbol{\nabla} \phi(\boldsymbol{r})$. 
To ensure accuracy in the phase computation, the phase must be unwrapped along both axes, denoted by $\phi^{'}_x$ and $\phi^{'}_y$.
The total velocity is then derived by combining the $x$ and $y$ components of the gradient, calculated from the unwrapped phase along both axes.

Direct access to the fluid phase, shown in Fig.\ref{fig:setup}(b), allows a measurement of the velocity field, given by $\boldsymbol{v^{tot}}(\textbf{r}) \propto \nabla_{\perp}\phi(\textbf{r})$ (see Supplementary for details).
We introduce the density-weighted velocity, given by $\boldsymbol{u^{tot}(r)}=\sqrt{\rho(\boldsymbol{r}})\boldsymbol{v}^{tot}(r)$, where $\rho(\boldsymbol{r})$ is the light intensity.
We then identify the compressible and incompressible parts of $\boldsymbol{u^{tot}(r)}$ using the Helmholtz decomposition to separate the divergent (compressible) and rotational (incompressible) components:
\begin{equation}
    \boldsymbol{u^{tot}}(\boldsymbol{r})=\underbrace{\boldsymbol{\nabla}\phi(\textbf{r})}_{\textrm{compressible}} + \underbrace{\boldsymbol{\nabla}\times\textbf{A}(\textbf{r})}_{\textrm{incompressible}}
\end{equation}
where $\phi$ is scalar and $\textbf{A}$ a vector field. 
The same decomposition can be written in the Fourier space:
\begin{equation}
    \boldsymbol{U^{tot}}(\boldsymbol{k}) = i\boldsymbol{k}U_\phi(\boldsymbol{k})+i\boldsymbol{k}\times\boldsymbol{U_{A}}(\boldsymbol{k}),
\end{equation}
where $U_\phi(\boldsymbol{k})=-i\frac{\boldsymbol{k}\cdot \boldsymbol{U^{tot}}(\boldsymbol{k})}{||\boldsymbol{k}||^2}$ and $\boldsymbol{U_{A}}(\boldsymbol{k})=i\frac{\boldsymbol{k}\times \boldsymbol{U^{tot}}(\boldsymbol{k})}{||\boldsymbol{k}||^2}$.\\
Thus, we  write the definition of the compressible and incompressible part in the real space:
\begin{equation}
    \begin{split}
        \boldsymbol{\nabla}\phi(\boldsymbol{r})&=\textrm{TF}^{-1}[i\boldsymbol{k}\cdot U_\phi(\boldsymbol{k})] \\
        \boldsymbol{\nabla}\times\textbf{A}(\boldsymbol{r})&=\textrm{TF}^{-1}[i\boldsymbol{k}\times \boldsymbol{U_A}(\boldsymbol{k})].
    \end{split}
\end{equation}
We obtain the incompressible weighted velocity field by directly subtracting the compressible part from the total weighted velocity $\boldsymbol{u^{inc}}=\boldsymbol{u^{tot}}- \boldsymbol{\nabla}\phi(\textbf{r})$.\\

\textit{Appendix B: Jones-Roberts Soliton velocity.}
In the localized soliton regime, the velocity $v$ in $c_s$ unit is obtained by fitting the phase jump profile with the analytical formula for a planar dark soliton \cite{tsuzuki_nonlinear_1971}:
\begin{equation}
    \Psi_s(x) = \sqrt{\rho}\left[ \sqrt{1-\frac{v^2}{c_s^2}} \textrm{tanh}\left( \frac{x}{\xi\sqrt{2}}\sqrt{1-\frac{v^2}{c_s^2}} \right) + i\frac{v}{c_s} \right].
    \label{soliton}
\end{equation}
Given that the phase returns to zero at greater distances, the discussion regarding the use of the planar soliton formula remains open. 
Depending on the size of the chosen region, the value of $v/c_s$ obtained from the fit will vary.
Here, we opted for a trade-off between spatial resolution and distance in units of $\xi$, stopping when the relative phase difference between the left and right sides of the soliton decreases by half. 
Consequently, all datas of Fig.\ref{fig:velocity} are processed within a $15\xi \times 15\xi$ window. 

Another approach to extract the velocity involves using a modified version of the asymptotic formula from \cite{tsuchiya_solitons_2008, krause2024thermal}:
\begin{equation}
    \phi(\mathbf{r}) = -2 \sqrt{2} \epsilon \frac{\epsilon x / \xi}{3/2 + \epsilon^4 y^2 / \xi^2 + \epsilon^2 x^2 / \xi^2}
\end{equation}
adapted for lower velocities by applying a Shanks transformation:  
$U_{\text{Shanks}} = U - \frac{(U - 1)^2}{U - 2}$ with $U=v/c_s$ (see Supplementary for details). Note that here we use the more exact $\epsilon\equiv\sqrt{2}\sqrt{1-v/c}$~\cite{jones_motions_1986}, which differs from the asymptotic form used in \cite{tsuchiya_solitons_2008} at smaller $v$, but yields more accurate estimate of the velocity.\\

\textit{Appendix C: Analytical results.} The standard approach for vortex spectra~\cite{bradley2012energy} leads to a spectrum that can't be analytically inverted to get spatial correlations. We introduce a new vortex ansatz that simplifies spectra enabling an analytic correlation function to be obtained in closed form~\footnote{The full calculation will be presented elsewhere; here we summarize the key results for our purpose.}. The vortex wavefunction is described using an exponential with the correct near and far field asymptotics, with the form
\begin{align}
    \psi_e(\mathbf{r}) &= \sqrt{n_0}(1-e^{-\Lambda r/\xi})e^{\pm i\theta},
\end{align}
where $\Lambda=0.8249...$ is the numerically determined slope at the core~\cite{bradley_spectral_2022}. Evaluating the Fourier transforms for the weighted velocity $\mathbf{u}=\sqrt{\rho}(v_x,v_y)$, we evaluate the incompressible velocity power spectrum (a single vortex in a homogeneous background is purely incompressible)
\begin{align}
    E^i(k)&\equiv \frac{m}{2}{k}\int_0^{2\pi}d\phi |\tilde{\mathbf{u}}(\mathbf{k})|^2.
\end{align}
For a single vortex the exponential ansatz gives $E^i_v(k)=(\pi\hbar^2n_0\xi/m) F(k\xi)$ where \begin{align}    
F(z)&=\frac{\Lambda^2}{(z^2+\Lambda^2)z}=\frac{1}{z}- \frac{z}{z^2+\Lambda^2}
\end{align}
has the expected $z^{-3}$ and $z^{-1}$ asymptotics for large and small $z$ respectively~\cite{bradley_spectral_2022}. This function captures the key spectral features of a vortex, and contains an infrared divergence stemming from the long-range vortex velocity field. Inverting the spectrum to position space, we can obtain velocity two-point correlations averaged over all angles by evaluating the correlation integral~\cite{bradley_spectral_2022}
\begin{align}
    G^i(r)&=\int_0^\infty E^i(k)J_0(kr)dk.
\end{align}
To do this analytically, we introduce an infrared regularising factor by making the replacement $F(z)\to F_\Gamma(z)$ where
\begin{align}
     F_\Gamma(z)&\equiv\frac{1}{\sqrt{z^2+\Gamma^2}}- \frac{z}{z^2+\Lambda^2},
\end{align}
and $\Gamma = 2\pi/R$ is an infrared cutoff at the dimensionless system scale $R$.
Each term can be integrated against $J_0(kr)$, giving the normalised two-point velocity correlation function $g^i(r)\equiv G^i(r)/G^i(0)$ for a single vortex 
\begin{align}
    g^i_v(r)&=\frac{K_0\left(\frac{\Gamma r}{2\xi}\right)I_0\left(\frac{\Gamma r}{2\xi}\right)-K_0\left(\frac{\Lambda r}{\xi}\right)}{\ln\left(\frac{2\Lambda}{\Gamma}\right)},
\end{align}
shown in Fig.~\ref{fig:corr}(a), using the infared cutoff corresponding to the quantum fluid radius $R=63\xi$. It agrees with the numerical result until the curvature of the background density becomes significant. Some further analysis gives an approximate treatment of the dipole and pair separated by distance $d$ in terms of the single vortex result in the form
\begin{align}
    g_d^i(r) &=\begin{cases}\dfrac{g_v^i(r)-g_v^i(d)}{1-g_v^i(d)}, &r\leq d \\
        0, &r>d.
    \end{cases}
     \label{gd}
\end{align}
and 
\begin{align}
    g_p^i(r) &=\begin{cases} \dfrac{g_v^i(r)+g_v^i(d)}{1+g_v^i(d)},&r\leq d\\
        \dfrac{2g_v^i(r)}{1+g_v^i(d)},&r>d
        \end{cases}
        \label{gp}
\end{align}
respectively. This approximate treatment neglects small compressible effects near the cores, but has the advantage of clearly showing the linear velocity cancellation and reinforcement effects: for $r>d$ the dipole has vanishing angle-averaged velocity correlation, while in the same region the pair acquires reinforcement of the single vortex correlation by a factor of 2. 

For a fast moving JRS ($v>0.85c$) there is an analytic wavefunction~\cite{tsuchiya_solitons_2008}. Retaining the dominant anisotropy in the phase due to the jump across the JRS, while neglecting the weaker elliptical envelope, we find an expression for the compressible velocity power spectrum, and use Eq.~(\ref{cor_spec}) find the angle-averaged normalized velocity correlatotion function $g^c_j(r)=f(r/\xi)$, where
\begin{align}
    f(x)&=\frac{2}{x^3(x^2+\lambda^2)^2}\Big[2x^3-x\lambda^2\notag\\
&+\frac{\lambda^2(\lambda^2+4x^2)}{\sqrt{x^2+\lambda^2}}\cosh^{-1}\left(\frac{x}{\lambda}\right)\Big],
\label{asymp}
\end{align}
where $\lambda\equiv\sqrt{3}\xi/\epsilon$ is the dimensionless scale setting the correlation length. While the analytic wavefunction is undefined for $v<0.85$, $g^c_j(r)$ has no pathologies and can be analytically continued to lower velocities. 

\end{document}


\preprint{}

\title{Supplementary Material for:\\
Observation of Jones-Roberts solitons in a paraxial quantum fluid of light}

\date{\today}


\maketitle


\section*{Experiment details}
\begin{figure}[h]
    \centering
    \includegraphics[width=0.7\columnwidth]{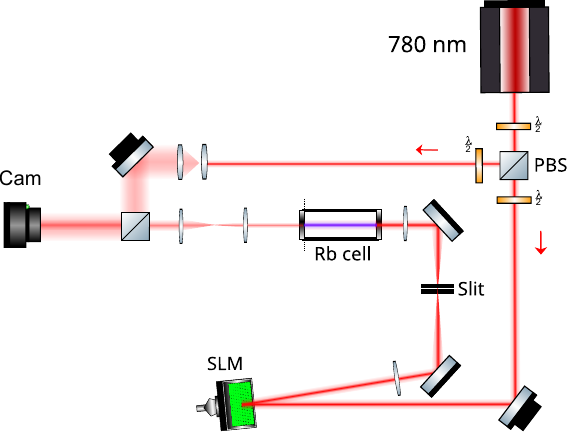}
    \caption{\textbf{Experimental Setup Detail} - A 780 nm laser beam is directed onto a Spatial Light Modulator (SLM), which is imaged at the input of a 20 cm-long rubidium vapor cell. The phase modulation produces two counter-propagating beams in the transverse plane. The output plane of the nonlinear medium is then imaged on a camera. A reference beam, separated from the initial laser beam, is recombined with the main beam before the camera to enable phase measurement.}
    \label{fig:manip_detail}
\end{figure}

To create the fluid of light, we use a continuous-wave diode laser at 780 nm, with tunable detuning relative to the $^{87}$Rb D2 resonance line, as illustrated in Fig. \ref{fig:manip_detail}. 
A $\pm 2\pi$ phase circulation pattern is applied to the Spatial Light Modulator (SLM) to generate the dipole vortex.
To eliminate unmodulated reflections from the SLM, a vertical grating is superimposed on the horizontal one, and only the first vertical order is selected in the Fourier plane using a slit.
The image from the SLM is then relayed to the entrance of the nonlinear medium via a telescope, with the beam waist at the entrance measuring $\omega = 1.7$ mm.
The nonlinear medium consists of a 20 cm rubidium cell containing a pure state of $^{87}$Rb.
The output plane of the cell is imaged onto a CCD camera without protective glass. 
Using a reference beam separated from the main beam upstream, we access the phase of the two counter-propagating beams by analyzing the interference fringes formed after recombination before the camera.
To adjust the effective interaction time $\tau$, the laser power is controlled using a rotating $\lambda/2$ wave plate, with a maximum power of 5W. 
During this process, the cell temperature is maintained at 150°C, and the laser detuning is fixed at $\Delta = -10$ GHz relative to the $^{87}$Rb $F = 2 \to F^{'}$ transition.

\section*{Dipole vortex time evolution}

\begin{figure}[!b]
    \centering
    \includegraphics[width=0.8\columnwidth]{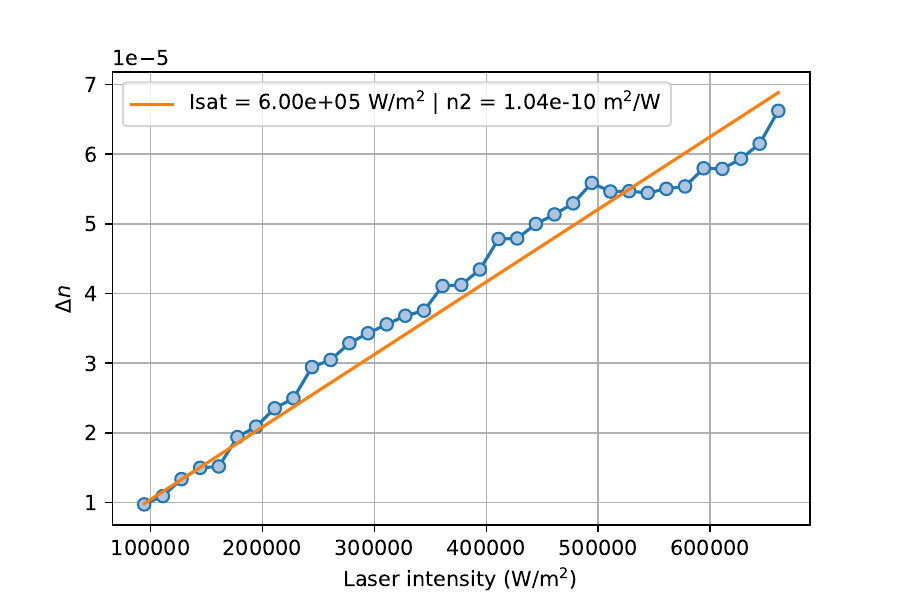}
    \caption{\textbf{Non-linear refractive index accumulated along the medium} - Each point is measured for a given beam intensity and the data are fitted from $n_2 I$, represented by the orange curve.}
    \label{fig:non-linear phase}
\end{figure}

To observe the time evolution of the dipole vortex, we vary the laser power and measure the fluid without the vortex (Gaussian profile) to determine the nonlinear phase shift $\Delta \phi$ accumulated within the cell. 
This phase shift is linked to the nonlinear refractive index through the formula $\Delta n = \Delta\phi / (k_0 L)$, as shown in Fig.\ref{fig:non-linear phase}.
This measurement allows us to obtain the value of $\tau = \Delta\phi$, and consequently, the parameter $\Bar{\xi} = \sqrt{L/(k_0 \tau)}$.
Next, we imprint the two phase circulations separated by a distance $\Delta \textbf{r}$ while maintaining the ratio $\Delta \textbf{r}/\Bar{\xi}$ constant.
The barycenter position of the two-phase circulation from the center of the beam $\textbf{r}_0$ is also adjusted in order to keep the quantity $\textbf{r}_0/\Bar{\xi}$ constant.

\begin{figure}[!t]
    \centering
    \includegraphics[width=0.7\columnwidth]{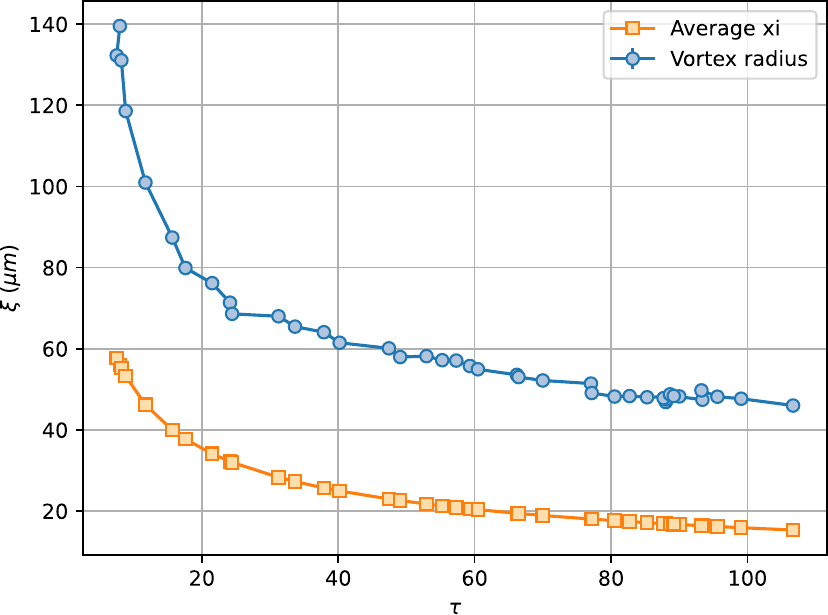}
    \caption{\textbf{Measurement of the final healing length} - Each point is measured for a given beam intensity and the data are fitted from .}
    \label{fig:xi_vortex}
\end{figure}

In addition to measuring the fluid with and without the dipole, we also measure the final value of $\xi$ for each $\tau$ shown in Fig.\ref{fig:xi_vortex} by injecting a single vortex and determining its radius through its radial amplitude profile.

\section*{Velocity map reconstruction}

\begin{figure}[h]
    \centering
    \includegraphics[width=0.9\columnwidth]{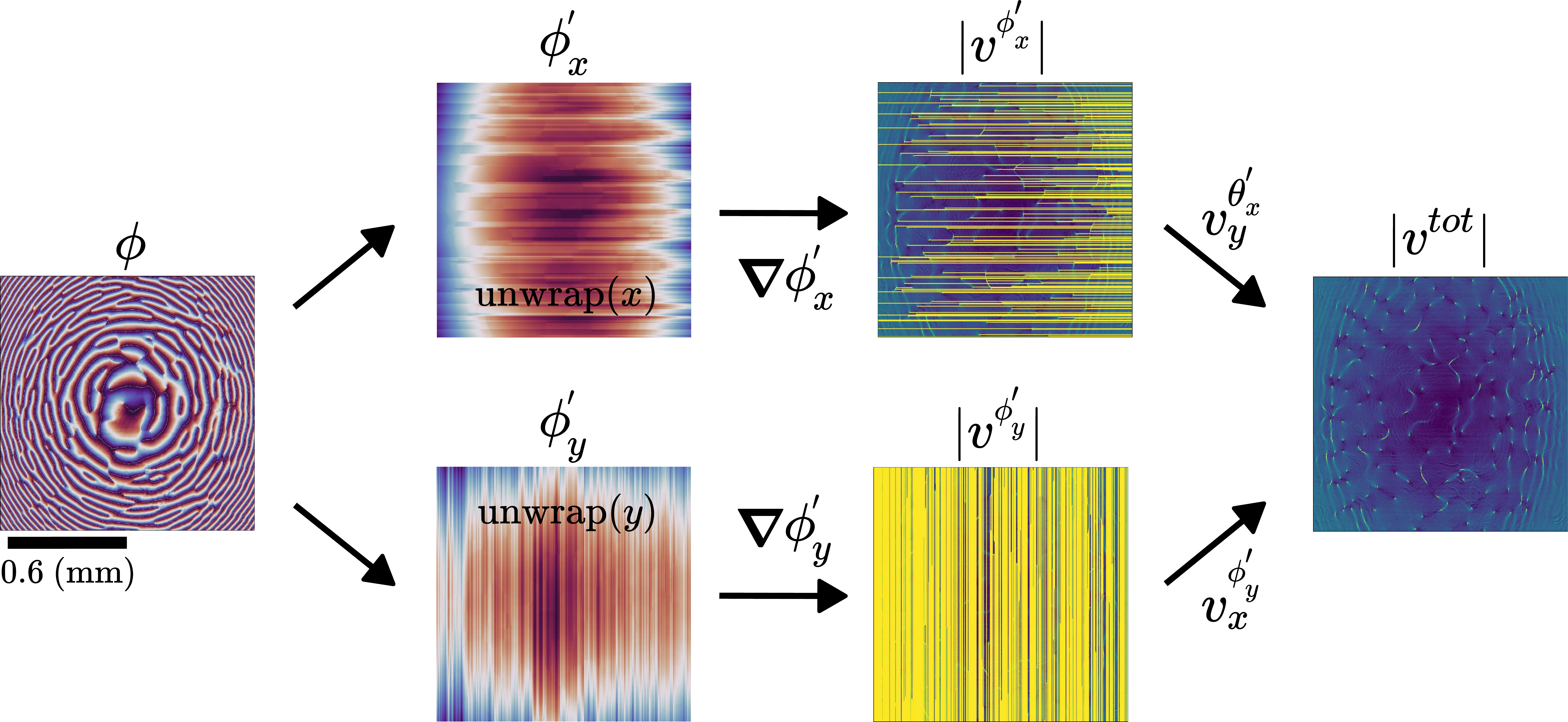}
    \caption{\textbf{Total velocity reconstruction} - The total velocity is the combination of each gradient component calculated from the 1D unwrapped phase.}
    \label{fig:velo_tot}
\end{figure}

From the phase map, we reconstruct the total velocity field, defined as $\boldsymbol{v}^{tot}(\boldsymbol{r}) = \boldsymbol{\nabla} \phi(\boldsymbol{r})$. 
To ensure accuracy in the phase computation, the phase must be unwrapped along both axes, denoted by $\phi^{'}_x$ and $\phi^{'}_y$.
The total velocity is then derived by combining the $x$ and $y$ components of the gradient, calculated from the unwrapped phase along both axes, as illustrated in Fig. \ref{fig:velo_tot}.

\section*{Shanks transformation}

\begin{figure}[!t]
    \centering
    \includegraphics[width=0.8\linewidth]{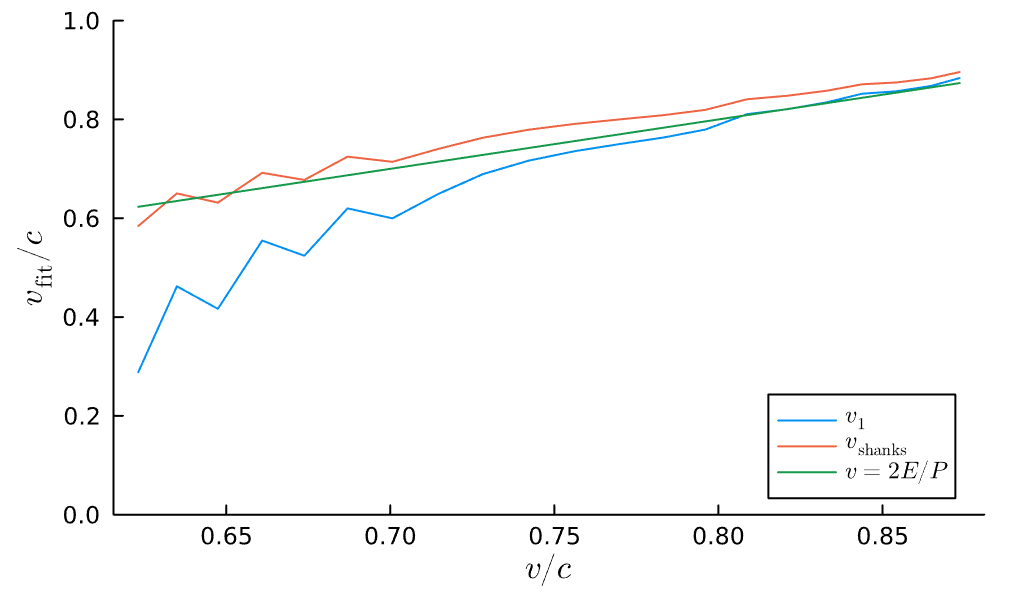}
    \caption{\textbf{Rarefaction pulse velocity comparison.} Exact velocity of the JRS (green curve) calculated using the formula $v = 2E/P$, where $E$ is the interaction energy and $P$ is the momentum. The fitted velocities $v_1$ and $v_{\text{Shanks}}$ are depicted by the blue and red curves, respectively.}
    \label{fig:shanks}
\end{figure}

To determine the speed of the Jones-Roberts soliton, we used several fitting methods, including one based on the asymptotic formula for the JRS phase described in the *End Matter*. 
However, since this formula is valid only for solitons with high velocities $v/c_s \geq 0.85$, we applied two transformations to extend its applicability.
\begin{itemize}
    \item Use $\epsilon = \sqrt{2} \sqrt{1 - v_1}$ as defined in the *End Matter*, where $v_1 = \frac{v}{c_s}$. 
This modification is equivalent to the asymptotic result $\sqrt{1-v_1^2}$ at high velocities but provides better agreement at lower speeds.
    \item Apply a Shanks transformation to $v_1$, resulting in the expression: $$v_{\text{Shanks}} = v_1 - \frac{(v_1 - 1)^2}{v_1 - 2}.$$
\end{itemize}
We tested these transformations against simulations of the Gross-Pitaevskii Equation, as shown in Fig.~\ref{fig:shanks}. 
In these simulations, the exact velocity is known using the formula $v = 2E/P$, where $E$ is the interaction energy and $P$ is the momentum (green curve). 
Also shown is the extracted values for $v_1$ and $v_{\text{Shanks}}$ from our fitting proceedure, depicted by the blue and red curves, respectively.
At low velocities $v_1$ deviates significantly from the true value given by the green curve. 
In contrast, the Shanks-transformed velocity remains close to the exact result across the entire range, demonstrating its robustness and accuracy.

\section*{Spectral analysis}
\begin{figure}[!t]
    \centering
    \includegraphics[width=1\columnwidth]{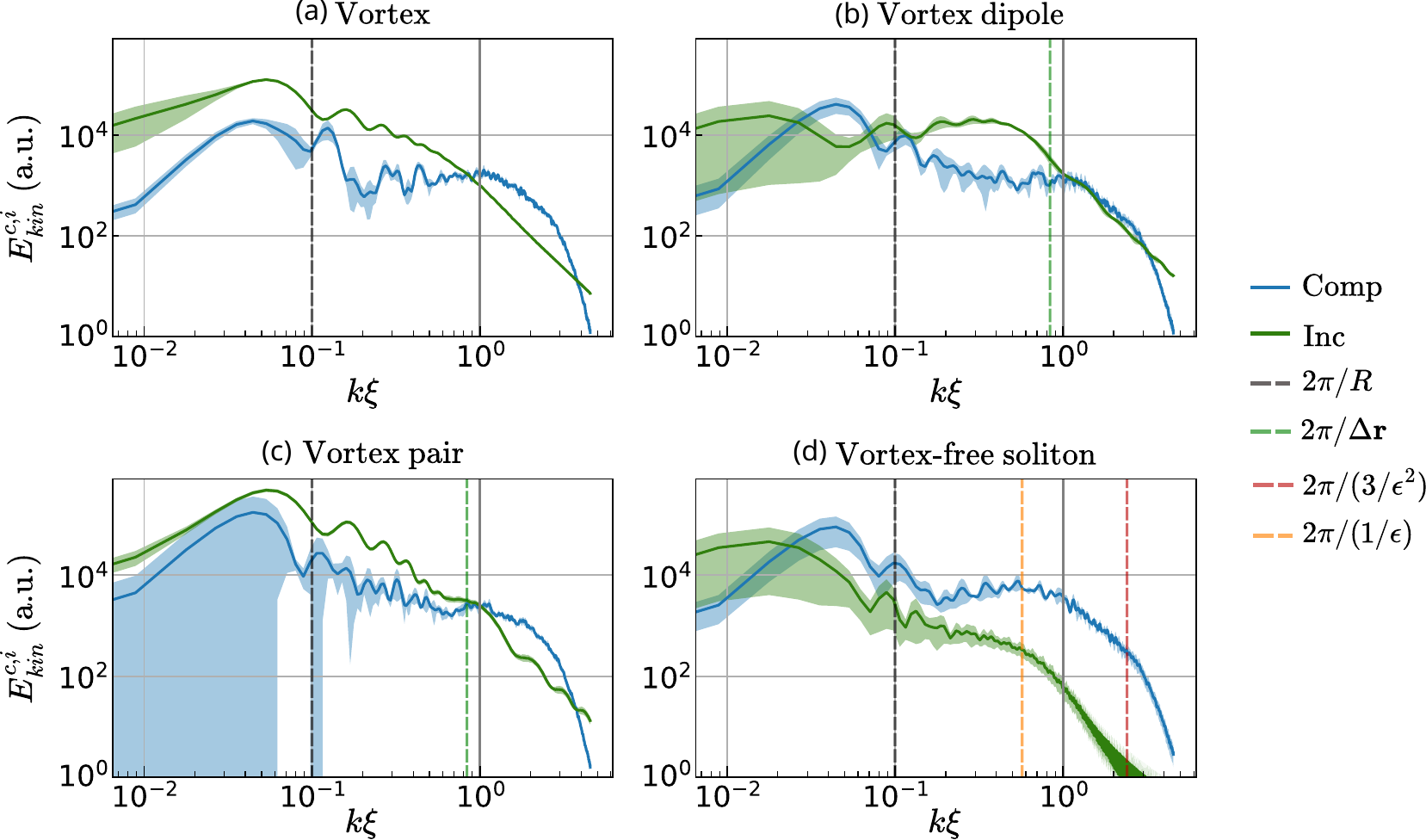}
    \caption{\textbf{Experimental kinetic energy spectra} - Kinetic energy spectra versus the adimensional unit $k\xi$ for a single vortex (a), a vortex dipole (b), a vorte pair (c), and a Jones-Roberts soliton (d).
    The compressible and incompressible part of the energy are respectively shown by the blue and green curves.}
    \label{fig:e_spectra}
\end{figure}

From the compressible and incompressible density-weighted velocity field, we compute the velocity power spectrum using the spectral analysis for compressible fluids [31].
The velocity power spectrum is given by
\begin{equation}
    E^{c,i}_{kin}(k) = \frac{m}{4\pi}\int d^2r kJ_0(k\textbf{r})C[\boldsymbol{u}^{c,i}, \boldsymbol{u}^{c,i}](\textbf{r})
    \label{e_spec}
\end{equation}
with $C[\boldsymbol{u}^{c,i}, \boldsymbol{u}^{c,i}](r)$ being the two-point correlation in position where $\boldsymbol{u}^{c,i}$ are the compressible and incompressible density-weighted velocity field and $J_0$ is the zero order Bessel function.\\
Fig.\ref{fig:e_spectra} shows the incompressible spectra (green curves) for a single vortex, a pair of same sign vortices and a vortex dipole. 
In the ultraviolet (UV) range ($k_\perp \xi > 1$), both spectra exhibit a $k^{-3}$ decay, which arises from the internal structure of the vortex core. 
In the infrared (IR) range ($k_\perp \xi \ll 1$), the spectra exhibit trends of $k^{-1}$ for the single/pair vortex and $k^{+1}$ for the vortex dipole, determined by whether the velocity field is rotational or irrotational, as described theoretically in [39].\\
The compressible contribution is negligible in these cases, given the total compressible-to-incompressible energy ratio of $0.3 \pm 0.1$ for the single vortex, $0.3 \pm 0.2$ for the vortex pair and $0.4 \pm 0.3$ for the vortex dipole. In the case of the vortex-free soliton, the compressible component (blue curve) dominates, with a comp/inc ratio of $5.2 \pm 0.6$, consistent with the absence of vortices. 
